\documentclass{article}
\usepackage{ijcai18}

\usepackage{times}
\usepackage{xcolor}
\usepackage{soul}
\usepackage[utf8]{inputenc}
\usepackage[small]{caption}
\usepackage{amsmath, amssymb} %aligned equations
\newcommand{\argmax}[1]{\underset{#1}{\operatorname{arg}\,\operatorname{max}}\;} %argmax

\newtheorem{assumption}{Assumption}[section]
\newtheorem{lemma}{Lemma}[section]
\newtheorem{algorithm}{Algorithm}[section]

\usepackage{bm}
\usepackage{mathtools}

\usepackage{pslatex} %Times font

\usepackage{booktabs} %tables
\usepackage[pdfborder={0 0 0}]{hyperref}%for hyperlinks without ugly boxes
\usepackage{graphicx} %images

\usepackage{caption}%custom size captions
\author{Irena Gao\\ 
Illinois Mathematics and Science Academy (IMSA)  \\
igao@imsa.edu}

\title{Fair Matching in Dynamic Kidney Exchange}
\begin{document}
\maketitle %title heading issues (makedate / space)

%%%%%%

\begin{abstract}
Kidney transplants are sharply overdemanded in the United States. A recent innovation to address organ shortages is a kidney exchange, in which willing but medically incompatible patient-donor pairs swap donors so that two successful transplants occur. Proposed rules for matching such pairs include static \textit{fair} matching rules, which improve matching for a particular group, such as highly-sensitized patients. However, in dynamic environments, it seems intuitively fair to prioritize time-critical pairs.  We consider the tradeoff between established sensitization fairness and time fairness in dynamic environments. We design two algorithms, SENS and TIME, and study their patient loss. We show that the there is a theoretical advantage to prioritizing time-critical patients (around 9.18\% tradeoff on U.S. data) rather than sensitized patients. Our results suggest that time fairness needs to be considered in kidney exchange. We then propose a batching algorithm for current branch-and-price solvers that balances both fairness needs.
\end{abstract}

\section{Introduction}
	Kidney disease has consistently been a leading health problem in the United States; in 2016, there were 44,193 nephritis mortalities, making it the 9th leading cause of death. In the general American population, poor kidney health is incredibly prevalent; an estimated 31 million people, or 10\% of the adult population, have chronic kidney disease (CKD), which may progress to end-stage renal disease (ESRD) or high risk for outright kidney failure.

\par

	It is vital that ESRD patients are promptly treated with either of two options: dialysis or transplantation. A transplant is significantly more affordable (\$32,000 followed by \$25,000 annually vs. dialysis's \$89,000 annually) and effective (80\% five-year survival rate vs. dialysis's 35\%) \cite{noauthor_statistics_nodate}; however, there has long been a severe---and increasing---organ supply shortage. In 2016, 54,261 patients were added to the kidney transplant waiting list, but only 19,060 transplants were performed with median wait times of well over two years \cite{noauthor_national_2017}.

	\par

	While there is a definite lack of donors, the kidney shortage is exacerbated by patients who may have willing, but medically incompatible live donors---who then must then join a kidney waitlist. To safely transplant, patients and potential donors must show positive \textit{crossmatch} (compatibility) in blood type. However, even if their donor's blood type matches, patients may be further limited by their \textit{tissue sensitization}, or probability of tissue crossmatch failure, as measured by a Calculated Panel Reactive Antibodies (CPRA) score between 0 (0\% probability of tissue crossmatch failure) and 100 (100\% probability of crossmatch failure). In our work, we refer to sensitization scores between $[0, 1]$, or the CRPA score divided by 100. Highly sensitized patients (CPRA score $\ge$ 80) are at a much higher risk of death, currently waiting three times longer for transplantation \cite{dickerson_price_2014}.

	\par

	A promising innovation to address the patient-donor mismatch problem---and the general kidney shortage---is a \textit{kidney exchange}. A kidney exchange allows medically incompatible patients to swap donors with other such patients, resulting in \textit{two} successful transplantations where there would have been none. Swaps can be generalized as \textit{k-cycles}, where $k$ patients give their donors to each other, resulting in $k$ transplantations. An altruistic donor, with no attached patient, can also start a \textit{chain}, where patients pay forward their donor beginning with the altruistic donor's recipient. Because no currency is involved, organ exchanges do not violate the National Organ Transplant Act of 1984 \cite{abecassis_consensus_2000}. Kidney exchanges present an interesting problem, approachable from both the economics and theoretical computer science angles: given a set of incompatible patient-donor pairs, how might we solve for the "optimal" set of cycles and chains?  

	\par

	Before this NP-hard problem can be approximated, the first step for the field is to articulate the problem; i.e. define, justify, and theoretically solve for "optimality" \cite{abraham_clearing_2007}. The standard definition is the earliest; it focuses on maximizing the number of patients matched at the time the problem is considered \cite{roth_kidney_2004}. This definition of "optimality" as maximizing the cardinality of patients matched is \textit{utilitarian}, while its simple consideration of only the present time period is called \textit{myopic}. Later work proposed \textit{dynamic} utilitarian methods, which emphasize maximizing the number of patients matched over an extended or infinite time period \cite{dickerson_dynamic_2012}\cite{unver_dynamic_2010}. Considering multiple time periods allows the central clearinghouse to strategically save valuable pairs for a later time period, where they might result in more overall matchings than if matched immediately. Ashlagi and Roth \shortcite{ashlagi_individual_2011} and Hajaj, et al. \shortcite{hajaj_strategy-proof_2015} also consider \textit{strategyproof} utilitarian methods, which ensure hospitals have no incentives to misreport their mismatched pairs to maximize personal gain at the cost of overall utilitarianism.

	\par

	Recently, several authors have also considered the question of \textit{fairness}. Dickerson \shortcite{dickerson_price_2014} defines a \textit{sensitization fairness} as prioritizing highly-sensitized patients, who, by definition, are harder to match and may remain unmatched while easier-to-match pairs are cleared. Dickerson quantifies the theoretical loss in matched patient cardinality compared to the myopic utilitarian algorithm as $\frac{2}{33}$ using United States averages. Separately, work by \cite{kahng_timing_2016}, \cite{anderson_dynamic_2014}, and \cite{akbarpour_dynamic_2014} raise the question of \textit{time fairness}, a natural concern in emerging dynamic problem formulations, which minimizes wait time for patients and may prioritize critically ill patients with higher mortality rates due to limited lifespan. As patients are diagnosed with ESRD at varying eGFR levels, we consider a pre-dialysis eGFR level of 5 mL/min/1.73 $m^2$/year as critical. Intuitively, it seems that strict sensitization fairness may come at the cost of time fairness, and vice versa.  Especially as kidney exchange models shift from static to dynamic and time fairness becomes an increasing concern, the kidney exchange clearing problem faces a major question: how should pairs be prioritized, and at what cost?

	\par 

%big claims here

	We explore the cost of prioritizing highly-sensitized patients over time-sensitive patients. Theoretically, we show that there is an advantage to prioritizing time-critical patients over highly-sensitized patients on the standard kidney exchange model. We then quantify our results using United State data, and finally, we propose a mechanism to consider both highly-sensitized and time-critical patients in sparse graphs.

\section{Model}
\subsection{Kidney Exchange}
	We consider the standard discrete-time kidney exchange model as established in \cite{dickerson_price_2014}, \cite{ashlagi_individual_2011}, \cite{roth_kidney_2004}, and \cite{unver_dynamic_2010} using the ABO blood model and two levels of sensitization. The kidney exchange pool at time $t$ is a set $V_t$ of patient-donor pairs $v = (p,d)$. Pairs enter the market at some time $t \ge 0$; the average quantity of incoming pairs per time step is $n$. Pairs leave the pool either when matched with another pair $w \in V_t$ or when $v$ perishes. A pair is denoted as \textit{critical} if its patient has an eGFR $\le$ 5 mL/min/1.73 $m^2$/year. Critical pairs have a \textit{perishing rate} of $\rho_C$ and non-critical pairs have a perishing rate of $\rho_{NC}$, where $\rho_{C} \gg \rho_{NC}$. A non-critical pair that is not matched has probability $\eta_C$ of becoming critical the next time step.

\par

	There are 32 \textit{types} of pairs, categorized by the CPRA score $p_s$ of $p$ and the blood types $X$ and $Y$ of $p$ and $d$, respectively. Begin by partitioning $V_t$ by blood type into 16 subsets $V_{X-Y}$. Next, as we consider only two levels of sensitization, consider a sensitization thershold dividing highly- and lowly-sensitized patient, $\sigma \in [0,1]$; in practice, $\sigma = 0.8$. Highly-sensitized pairs have sensitization level $\gamma_H$; lowly-sensitized pairs have $\gamma_L$. The average sensitization level $\bar{\gamma} = \mu_H \cdot  \gamma_H + (1-\mu_H) \cdot \gamma_L$. Further partition each subset $V_{X-Y}$ into $\{V^L_{X-Y} \cup V^H_{X-Y}\}$, such that:
\begin{itemize}
	\item $V^L_{X-Y}$ is of patient-donor blood types $X-Y$ and not highly-sensitized: $\{v | v \in V_{X-Y}, p_s < \eta\}$
	\item $V^H_{X-Y}$ is of patient-donor blood types $X-Y$ and highly-sensitized:  $\{v | v \in V_{X-Y}, p_s \ge \eta\}$
\end{itemize}
A future pair incoming at some time $\tau > t$ has an \textit{arrival probability} $\mu_{H}$ of being highly-sensitized and arrival probability $1 - \mu_{H}$ of being not highly-sensitized. The average proportion of the entering pool that is critical is $\mu_{C}$. The probability of an individual (patient or donor) arriving with blood type $X$ is $\mu_X$. Therefore, at $\tau$, 

\begin{align} %& sets alignment point
	\mathbf{E}(|V^H_{X-Y}|)&\le n \cdot \mu_{X} \cdot \mu_{Y} \cdot \mu_H
\\
	\mathbf{E}(|V^L_{X-Y}|)&\le n \cdot \mu_{X} \cdot \mu_{Y} \cdot (1-\mu_H)
\end{align}

\par

	$V_t$ induces a \textit{compatability graph} $G_t={V_t, E(V_t)}$, in which the nodes are pairs, and $v$ forms an edge with another pair $w \in V_t$ only if the donor of $w$ is compatible with the patient of $v$. A cycle $c$ in $G$ represents a kidney exchange where each vertex in $c$ obtains the kidney of the previous vertex. Cycles may be of length $k$. We limit our cycles to $k=3$, as Ashlagi and Roth \shortcite{ashlagi_individual_2011} prove that this length is sufficient for a perfect matching. For simplicity, we do not consider altruistic chains. A matching $m$ is a collection of vertex-disjoint cycles in $G$. The set of all legal matchings on G is $M$.

\subsection{Tradeoff Between Two Algorithms}
We study the tradeoff between prioritizing sensitization fairness over time fairness through two algorithms: SENS and TIME. Consider the set of all highly-sensitized pairs $H_t \subseteq V_t$ and the set of all critical pairs $C_t \subseteq V_t$.  SENS maximizes the highly-sensitized pairs matched, while TIME maximizes the critical pairs matched. Formally:

\begin{align}
	SENS(G_t) = \argmax{m \in M} |\{v \in m \land v \in H_t\}|
\\
	TIME(G_t) = \argmax{m \in M} |\{v \in m \land v \in C_t\}|
\end{align}

As these algorithms will always match their target pairs over another similar pair when possible, we argue that SENS and TIME upper bound the potential \textit{losses} due to pursuing each brand of fairness. An algorithm's loss at $t$ is defined as the number of agents who perish after the clearinghouse acts during $t$. Formally:

\begin{equation}
	L^{t}(ALG) = \rho_C  |\{v \notin m_{ALG} \land v \in C_t\}| + \rho_{NC} |\{v \notin m_{ALG} \land v \notin C_t\}|
\end{equation}

We study the loss, rather than the utility, of algorithms. We make two arguments for this approach. First, the very nature of time fairness, concern over critically ill patients with a higher perishing rate, prioritizes the question of \textit{how many might we lose?} rather than \textit{how many might we not match?} Second, choosing algorithms that minimize loss, rather than maximizing utility, may be fairer. Lost pairs are lost forever, but unmatched, unperished pairs may actually improve their matchability when new pairs enter the pool in later rounds. Since these pairs lose much less than pairs who perish during the time step, we study our algorithms through their losses.

\par

The tradeoff between SENS and TIME is defined similarly to Dickerson's price of fairness \shortcite{dickerson_price_2014}:

\begin{equation}
	TRDOFF(SENS, TIME) = \lim_{\tau \to \infty} \frac{L^{\tau}(SENS) - L^{\tau}(TIME)}{L^{\tau}(SENS)}
\end{equation}

Note that we study this tradeoff in a \textit{dynamic} environment (i.e. we consider the tradeoff as $t \to \infty$, rather than just at $t=0$); as of this writing, only \cite{ashlagi_kidney_2013} has also studied fairness in dynamic matching, and then only sensitization fairness.

\section{Theory}

In this section, we upper bound the tradeoff between SENS and TIME on dense graphs. We show that, with high probability, the theoretical tradeoff is upper bounded by 2.385. 

\subsection{Setup}

We begin by partitioning $V_t$ into $\{V^O \cup V^U \cup V^S \cup V^R\}$ based on the ease of finding another blood-compatible pair. We define a universal binary relation $\vartriangleleft$ over the four blood types. Blood type $X \vartriangleleft$ blood type $Y$ if $X$ can receive a kidney from donor of blood type $Y$.  We then create four subsets of $V_t$:

\begin{itemize}
\item $V^O = \{V_{X-Y} : X \vartriangleleft Y \land X \neq Y\}$ (i.e. the patient is offering a donor more valuble than the one they seek. These pairs may still be in $V_t$ due to tissue incompatability with probability $\bar{\gamma}$)
\item $V^U = \{V_{X-Y} : Y \vartriangleleft X \land X \neq Y\}$ (i.e. the patient is seeking a donor more valuble than the one they offer)
\item $V^S = \{V_{X-X}\}$ (i.e. the patient and donor have the same blood type)
\item $V^R = \{V_{X-Y} : Y \ntriangleleft X \land X \ntriangleleft Y\}$ (i.e. pairs A-B or B-A)
\end{itemize}

We make the same assumptions as Ashlagi and Roth \cite{ashlagi_individual_2011}, which are supported by United States data (see below).

\begin{assumption}[Average sensitization of incoming patient]
	 $\mu{H} < \bar{\gamma} < \frac{1}{2}$
	since, in practice, $\mu{H}$ is $\frac{1}{3}$ \cite{noauthor_highly_nodate} and $\gamma_H \in [0.8, 1.0)$.
\end{assumption}

\begin{assumption}[Blood type frequency]
	$\mu_O < 1.5\mu_A$ \cite{dickerson_price_2014}
\end{assumption}

Under these assumptions, Ashlagi and Roth's Lemma 5.1 proves that, with high probability in a dense graph, there exists a matching $M^*$ that matches all of $V^O$, $V^S$, and $V^R$, though not all of $V^U$ \shortcite{ashlagi_individual_2011}. We use this as the foundation for our work.

\subsection{Loss of SENS}
In this section, we solve for the loss of SENS for some time $\tau > 0$.
\begin{lemma}[Loss of SENS]
	\begin{multline*}
		L^{\tau}(SENS) \le \rho_C \bm{[} \sum_{k=0}^{\tau} \mu_C \cdot \mu_{O-AB}^{H} \cdot (1-\rho_C)^k \\
+ \eta_C \sum_{k=0}^{\tau} [ \sum_{j=0}^{k} (1-\mu_C) \cdot \mu_{O-AB}^{H} \cdot (1-\rho_{NC})^j ] (1-\rho_C)^{\tau - k} \bm{]} \\
+ \rho_{NC} \bm{[} \sum_{k=0}^{\tau} (1-\mu_C) \cdot \mu_{O-AB}^{H} \cdot (1-\rho_{NC})^k \bm{]}
	\end{multline*}
\end{lemma}
\subsubsection{for $t=0$}
	We first consider the SENS algorithm's loss on a newly entered pool at t=0; i.e. there are no pairs in $V_0$ before $n$ pairs are added to $V_0$.

\par

SENS seeks to match the maximum cardinality of matched and highly sensitized pairs. By Ashlagi and Roth's Lemma 5.1, all pairs in $V^O$, $V^S$, and $V^R$ are matched; trivially, this includes highly-sensitized pairs. Therefore, the only remaining unmatched highly-sensitized pairs are in $V^U = \{ V_{A-AB}, V_{B-AB}, \allowbreak V_{O-A}, V_{O-B}, V_{O-AB}\}$. We then exhaustively consider how many pairs in $V^U$ remain unmatched, as in Dickerson's proof \shortcite{dickerson_price_2014}. 

\par

As many B-AB pairs as possible have been matched in 2-cycles to AB-B pairs. Note that $|V^{H}_{B-AB}| \le n \cdot \mu_{B} \cdot \mu_{AB} \cdot \mu_{H}$, while $|V_{AB-B}| \le n \cdot \mu_{B} \cdot \mu_{AB} \cdot \bar{\gamma}$. Under assumption 3.1, $|V_{AB-B}| \ge |V^{H}_{B-AB}|$. Therefore, all $v \in V^{H}_{B-AB}$ are matched. Similarly, all $v \in V^{H}_{A-AB}$ are also matched. Any AB-O pairs intended to be matched in a 3-cycle with O-A and A-AB can still be matched in a 2-cycle with e.g. O-A at no loss.

\par

We used 2-cycles to match $V^{H}_{O-A}$ to $V_{A-O}$ and $V^{H}_{O-B}$ to $V_{B-O}$. Also by assumption 3.1, all of $V^{H}_{O-A}$ and $V^{H}_{O-B}$ will be matched. However, these 2-cycles may exhaust either the $V_{A-O}$ or $V_{B-O}$ pairs used to match $V_{A-B} - V_{B-A}$ or $V_{B-A} - V_{A-B}$ in Step 3 of Ashlagi and Roth's Lemma 5.1, preventing up to $|V^{r}_{B-A or A-B}|$ lowly-sensitized pairs from being matched. By Lemma 5.1 of Ashlagi and Roth \shortcite{ashlagi_individual_2011}, this difference is $o(n)$, smaller than any subgroup and sublinear in $n$; therefore, this term is insignificant as $n \rightarrow \infty$.

\par

The only remaining group, $V_{O-AB}$, cannot be matched without replacing another pair: swapping out an AB-O, O-A, A-AB 3-cycle for an AB-O, O-AB 2-cycle results in loss of one pair, breaking up two AB-B, B-O (or AB-A, A-O) 2-cycles for an AB-B, B-O, O-AB (AB-A, A-O, O-AB) 3-cycle also results in loss of one pair. Therefore, the overall number of unmatched pairs $\le n \cdot \mu_{O} \cdot \mu_{AB} \cdot \mu_{H}$.

\begin{figure}[h] %bottom of page
  \centering
    \includegraphics[width=1.0\columnwidth]{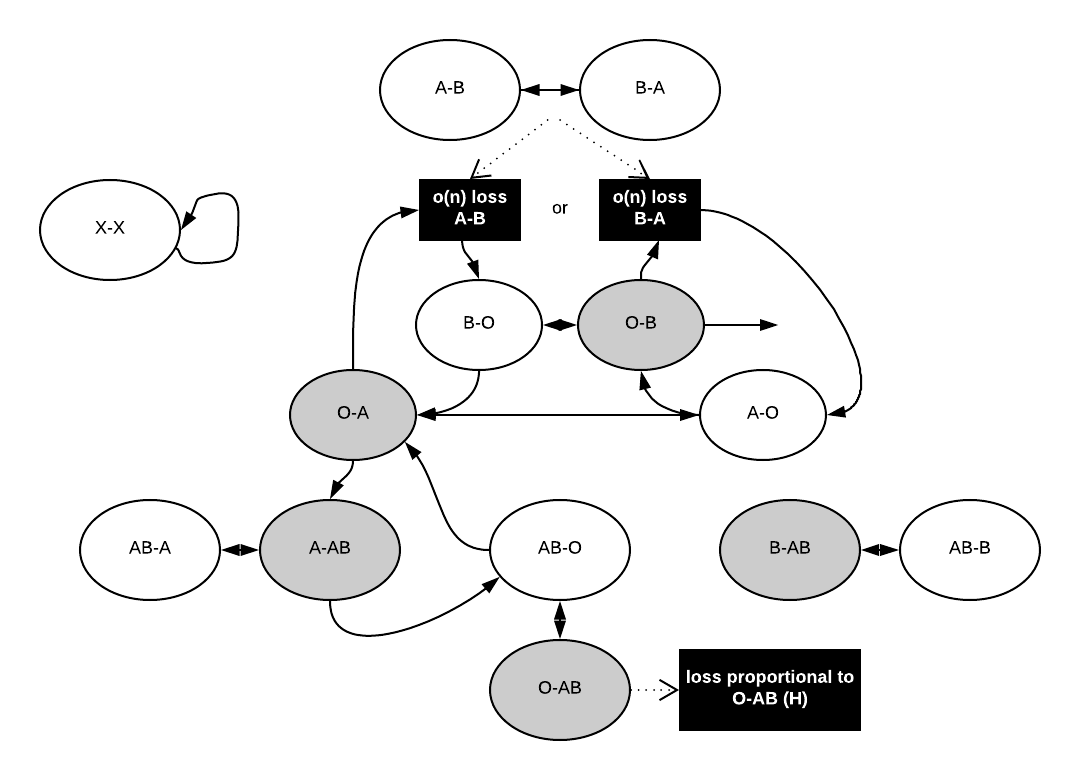}
  \caption{The matching proposed in Ashlagi and Roth's Lemma 5.1 is shown as the arrows between pair blood type nodes. Highly-sensitized pair blood types that we exhaustively consider above are shown with a gray background. Losses due to SENS's modifications have a black background.}
\end{figure}

\par

Let $\mu_{O-AB}^{H} = n \cdot \mu_{O} \cdot \mu_{AB} \cdot \mu_{H}$. Of these lowly-sensitized, unmatched pairs, $\mu_{O-AB}^{H} \cdot \mu_{C}$ are critical, while $\mu_{O-AB}^{H} \cdot (1-\mu_{C})$ are not. Therefore: 

\begin{equation}
	L^{0}(SENS)_{t=0} \le \rho_{C} \bm{[} \mu_{C} \cdot \mu_{O-AB}^{H} \bm{]} + \rho_{NC} \bm{[} (1-\mu_{C}) \cdot \mu_{O-AB}^{H} \bm{]}
\end{equation}

\subsubsection{for $t > 0$}
Next, we consider SENS's loss at $t=1$. There are $|V^{0}_{O-AB}| \le \mu_{O-AB}^{H}$ lowly-sensitized pairs already in the pool from $t=0$. The new pairs from $t=1$ are then added to the pool. After the above steps are run again, there are, at most, 

\[
\underbrace{\textstyle     \overbrace{\textstyle    \mu_{O-AB}^{H} \cdot \mu_{C} \cdot (1-\rho_C)}^{\mathclap{\text{unperished, critical}}} + \overbrace{\textstyle    \eta_C \cdot (1-\mu_C) \cdot \mu_{O-AB}^{H}}^{\mathclap{\text{non-critical who became critical}}}}_{\mathclap{t = 0}} + \underbrace{\textstyle      \mu_{O-AB}^{H} \cdot \mu_C}_{\mathclap{t=1}}
\]

critical pairs. Becoming critical and perishing are considered separate events, so the non-critical pairs who became critical are not multiplied by $(1-\rho_{NC})$. There are also, at most,

\[
\underbrace{\textstyle      (1-\mu_{C}) \cdot \mu_{O-AB}^{H} \cdot (1-\rho_{NC}) }_{\mathclap{t=0}} + \underbrace{\textstyle      (1-\mu_{C}) \cdot \mu_{O-AB}^{H} }_{\mathclap{t=1}}
\]

non-critical pairs. Therefore:

\begin{multline}
	L^{1}(SENS) \le \rho_{C} \bm{[} \mu_{O-AB}^{H} \cdot \mu_{C} \cdot (1-\rho_C) + \eta_C \cdot (1-\mu_C) \cdot \mu_{O-AB}^{H} \\
+ \mu_{O-AB}^{H} \cdot \mu_C \bm{]} + \rho_{NC} \bm{[}  (1-\mu_{C}) \cdot \mu_{O-AB}^{H} \cdot (1-\rho_{NC}) \\
+ (1-\mu_{C}) \cdot \mu_{O-AB}^{H} \bm{]}
\end{multline}

And for $t=2$, there are

\begin{multline*}
\underbrace{\textstyle    \mu_{O-AB}^{H} \cdot \mu_{C} \cdot (1-\rho_C)^2 + \eta_C \cdot \overbrace{\textstyle     (1-\mu_C) \cdot \mu_{O-AB}^{H} \cdot (1-\rho_C)}^{\mathclap{\text{non-critical pairs from } t=0 \text{, now critical at } t=1}}}_{\mathclap{t = 0}} \\ 
+ \underbrace{\textstyle    \mu_{O-AB}^{H} \cdot \mu_{C} \cdot (1-\rho_C)}_{\mathclap{t = 1}} \\
+ \underbrace{\textstyle    \eta_C  \overbrace{\textstyle      \bm{[} (1-\mu_C) \cdot \mu_{O-AB}^{H} \cdot (1-\rho_{NC}) + (1-\mu_C) \cdot  \mu_{O-AB}^{H} \bm{]}}}^{\mathclap{\text{non-critical pairs from } t=1 \text{ (including non-critical, untransformed pairs from t=0)}}}_{\mathclap{t = 1}} \\ 
+  \underbrace{\textstyle     \mu_{C} \cdot \mu_{O-AB}^{H} }_{\mathclap{t=2}}
\end{multline*}

critical pairs and

\begin{multline*}
\underbrace{\textstyle      (1-\mu_{C}) \cdot \mu_{O-AB}^{H} \cdot (1-\rho_{NC})^2 }_{\mathclap{t=0}} \\
+ \underbrace{\textstyle      (1-\mu_{C}) \cdot \mu_{O-AB}^{H} \cdot (1-\rho_{NC})}_{\mathclap{t=1}} \\
+  \underbrace{\textstyle      (1-\mu_{C}) \cdot \mu_{O-AB}^{H} }_{\mathclap{t=2}}
\end{multline*}

non-critical pairs. In general, for $t=\tau$ and $\tau > 0$, there are 

\begin{multline*}
\sum_{k=0}^{\tau} \mu_C \cdot \mu_{O-AB}^{H} \cdot (1-\rho_C)^k \\
+ \eta_C \sum_{k=0}^{\tau} [ \sum_{j=0}^{k} (1-\mu_C) \cdot \mu_{O-AB}^{H} \cdot (1-\rho_{NC})^j ] (1-\rho_C)^{\tau - k}
\end{multline*}

critical pairs and

\[
\sum_{k=0}^{\tau} (1-\mu_C) \cdot \mu_{O-AB}^{H} \cdot (1-\rho_{NC})^k
\]

non-critical pairs, making the total loss for $\tau > 0$

\begin{multline}
	L^{\tau}(SENS) \le \rho_C \bm{[} \sum_{k=0}^{\tau} \mu_C \cdot \mu_{O-AB}^{H} \cdot (1-\rho_C)^k \\
+ \eta_C \sum_{k=0}^{\tau} [ \sum_{j=0}^{k} (1-\mu_C) \cdot \mu_{O-AB}^{H} \cdot (1-\rho_{NC})^j ] (1-\rho_C)^{\tau - k} \bm{]} \\
 + \rho_{NC} \bm{[} \sum_{k=0}^{\tau} (1-\mu_C) \cdot \mu_{O-AB}^{H} \cdot (1-\rho_{NC})^k \bm{]}
\end{multline}

\subsection{Loss of TIME}
In this section, we solve for the loss of algorithm TIME for some time $\tau > 0$.
\begin{lemma}[Loss of TIME]
	\[
		L^{\tau}(TIME) \le \rho_{NC} \bm{[} \sum_{k=0}^{\tau} \mu_{O-AB}^{C} \cdot (1-\rho_{NC})^k \bm{]}
	\]
\end{lemma}
\subsubsection{for $t=0$}

We use a parallel method for algorithm TIME as we did for SENS. In the last step, the overall number of unmatched pairs $\le \mu_{O-AB}^{C}$, where $\mu_{O-AB}^{C} = n \cdot \mu_{O} \cdot \mu_{AB} \cdot \mu_{C}$; all are non-critical pairs. Therefore, the loss for TIME is:

\begin{equation}
	L^{0}(TIME) \le \rho_{NC} \bm{[} \mu_{O-AB}^{C} \bm{]}
\end{equation}

\subsubsection{for $t>0$}

For $t>0$, no critical pairs carry over from the previous time step; therefore, there are no critical pairs. However, there are 

\[
\sum_{k=0}^{\tau} \mu_{O-AB}^{C} \cdot (1-\rho_{NC})^k
\]

non-critical pairs. Therefore, for $t=\tau$:

\begin{equation}
	L^{\tau}(TIME) \le \rho_{NC} \bm{[} \sum_{k=0}^{\tau} \mu_{O-AB}^{C} \cdot (1-\rho_{NC})^k \bm{]}
\end{equation}

\subsection{Tradeoff between SENS and TIME}

We use our above results to calculate the tradeoff defined by equation (6). 

\begin{multline}
TRDOFF(SENS, TIME) = \\
\lim_{\tau \to \infty} \frac{
	\begin{aligned}
		\rho_C \bm{[} \sum_{k=0}^{\tau} \mu_C \cdot \mu_{O-AB}^{H} \cdot (1-\rho_C)^k 
		\\[-4\jot]
		+ \eta_C \sum_{k=0}^{\tau} [ \sum_{j=0}^{k} (1-\mu_C) \cdot \mu_{O-AB}^{H} \cdot (1-\rho_{NC})^j ] (1-\rho_C)^{\tau - k} \bm{]} 
		\\[-4\jot]
		+ \rho_{NC} \bm{[} \sum_{k=0}^{\tau} (1-\mu_C) \cdot \mu_{O-AB}^{H} \cdot (1-\rho_{NC})^k \bm{]} 
		\\[-4\jot]
		- \rho_{NC} \bm{[} \sum_{k=0}^{\tau} \mu_{O-AB}^{C} \cdot (1-\rho_{NC})^k \bm{]} 
	\end{aligned}
}{
	\begin{aligned}
		\rho_C \bm{[} \sum_{k=0}^{\tau} \mu_C \cdot \mu_{O-AB}^{H} \cdot (1-\rho_C)^k 
		\\[-4\jot]
		+ \eta_C \sum_{k=0}^{\tau} [ \sum_{j=0}^{k} (1-\mu_C) \cdot \mu_{O-AB}^{H} \cdot (1-\rho_{NC})^j ] (1-\rho_C)^{\tau - k} \bm{]}
		\\[-4\jot]
		+ \rho_{NC} \bm{[} \sum_{k=0}^{\tau} (1-\mu_C) \cdot \mu_{O-AB}^{H} \cdot (1-\rho_{NC})^k \bm{]}
	\end{aligned}
}
\end{multline}

In the United States, $\mu_{O} = 0.44$, $\mu_{AB} = 0.10$, $\mu_{H} = 0.3$ \cite{dickerson_price_2014}, $\mu_{C} = 0.14$ (where critical patients are those who begin treatment with eGFR $<$ 5 mL/min/1.73 $m^2$/year) \cite{noauthor_2016_nodate}, $\rho_{NC} = 0.25$, and $\eta_C = 0.09$ \cite{al-aly_rate_2010}. There is limited data about the relationship between pre-ESRD eGPR scores and mortality after dialysis begins; however, based on average mortality increases of 11\% based on a severe eGFR decline in the patient ($<$ -10 mL/min/1.73 $m^2$/year) by \cite{sumida_association_2016}, we set $\rho_{C} = 0.35$. Using these numbers, we estimate the tradeoff betwen SENS and TIME in the United States. However, since the double summation in $LOSS^{\tau}(SENS)$ is not computable, we estimate the tradeoff at $\tau=10$:

\[
TRDOFF(SENS, TIME) \approx \frac{0.010496n}{0.114249n} \approx 0.091872
\]

In other words, the long-term difference between prioritizing sensitized and prioritizing time-critical patients has an upper bound of around 9.18\% of SENS's unique loss; theoretically, matching sensitized patients over time-critical ones results in moderate losses. (We note, of course, that this value depends on the constants defined above, of which $\eta_C$ and $\rho_{NC}$ are still currently under research.) In real-world, sparser graphs, the loss is likely to be amplified, as Dickerson's Price of Fairness was amplified dramatically \cite{dickerson_price_2014}. In sparse graphs, it is much harder to find a matching that matches all of $V^O$, $V^S$, and $V^R$, so losses for both functions would likely increase. However, we informally argue that TIME would maintain its advantage over SENS, as it, by nature, attempts to minimize chances for losses. But at the very least, we prove that time fairness is theoretically advantageous to sensitization fairness.

\section{An Algorithm for Real-World Graphs}
Given our results, we propose an algorithm for sparse graphs that emphasizes matching critical patients while balancing both sensitization- and time-based fairness. This algorithm simply breaks up the traditional one-batch solution into four separate runs, so it is simply layered on top of current (or future) kidney exchange solvers. The technique of layering an algorithm on top of current solvers has been used by \cite{dickerson_dynamic_2012}, although their algorithm is completely different (a parameter-learning weighting system for dynamic matching).

\begin{algorithm}[A batching algorithm for both fairnesses]
\item %to add a space because par doesn't work
\begin{enumerate}
\item The solver considers only pairs in $C_t$ and matches them to each other. Unmatched pairs are passed down to the pool in Step 2.
\item Leftover pairs from $C_t$ are matched to pairs in $H_t$. Unmatched pairs in $C_t$ are passed down to the pool in Step 4; unmatched pairs in $H_t$ are passed down to pairs in Step 3.
\item Pairs in $H_t$ are matched to other pairs in $H_t$. Leftover pairs are passed down to Step 4.
\item All remaining pairs are matched.
\end{enumerate}
\end{algorithm}

\begin{figure}[h] %bottom of page
  \centering
    \includegraphics[width=1.0\columnwidth]{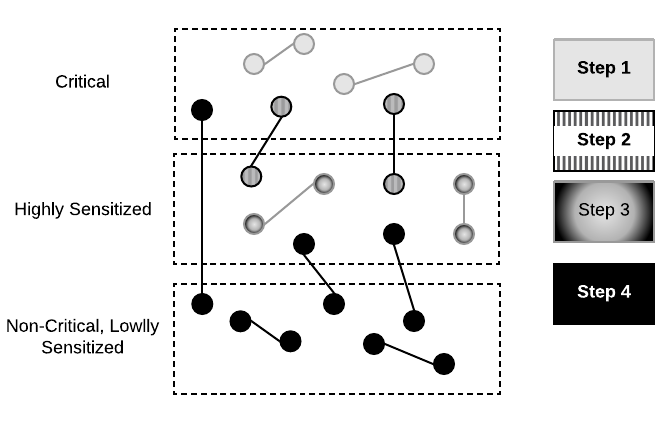}
  \caption{Illustration of Algorithm 4.1}
\end{figure}

In addition to relieving stress on the branch-and-price solver by addressing smaller groups at a time, this algorithm also avoids a computationally-heavy weighting function---valid in the future, but not for immediate implementation.

\section{Conclusion}

Before kidney exchanges can run, they must decide on a matching rule. Where Dickerson \shortcite{dickerson_price_2014} showed that utilitarian matching rules undermine sensitized patients in static settings, we theoretically showed that sensitization-based matching rules result in patient losses in dynamic settings. We also present a new definition of fairness, lexicographical time fairness, which maximizes the matching of critically-ill, time-sensitive pairs, and we propose a batched algorithm to balance time and sensitization needs. We contribute by being one of the first to consider dynamics fairness, uniquely evaluate fairness using losses, and the first to propose a time fairness. 

\par

Our work suggests that time plays a larger role in kidney exchange than currently acknowledged. Minimizing wait times and quickly clearing critical patients likely requires more than a modification to current branch-and-price solvers; they instead require an explicitly dynamic solver. We essentially justify recent attempts to entirely reformulate static solvers into dynamic ones, especially theoretical work in the formal Markov Decision Process environment \cite{akbarpour_dynamic_2014} \cite{anderson_dynamic_2014}. These solvers naturally optimize time fairness by weighting matches based on long-term utility with future-discounting; only sensitization fairness needs to be later imposed. 

\par

In the end, it is important to remember that kidney problems affect over a tenth of the United States population, making kidney exchange an important, and so time-pressed, innovation. Yet the care we take to design matching rules is more than theorizing; they affect the fairness, efficiency, and health of us all.

%%%%%%

\clearpage

\bibliography{2017-09-14_bib}
\bibliographystyle{named}

\end{document}